\documentstyle[preprint,aps]{revtex}
\begin{document}
\title{DARK MATTER AND THE VOIDS}
\author{ MOFAZZAL AZAM}
\address{THEORETICAL PHYSICS DIVISION, CENTRAL COMPLEX,
BHABHA ATOMIC RESEARCH CENTRE \\
TROMBAY, MUMBAI-400085, INDIA}
\maketitle
\vskip .5 in
\begin{abstract}
It is now, generally, believed that the presence of some form of
dark matter is essential to explain the flat rotation curves of
galaxies, and anomalous large velocities of galaxies  in the
clusters and superclusters. This dark matter  turns out to be
too elusive and , so far, there is no direct evidence for such
matter although there is a large variety of candidates proposed by
theoretical models. One would expect that the existence of vast
amount of dark matter would lead to the formation of some dark
galaxies or clusters which could be seen by gravitational lensing
events.This could serve as a direct evidence for dark matter.
However, so far, there is no conclusive lensing event pertaining to that.
\par In spite of absence of any direct evidence,
there is a consensus among physicists in favour of some form of
non-baryonic dark matter. In this paper, we take a different viewpoint
and would like to view dark matter as an {\bf emergent} phenomenon -
very much like many such phenomena known in condensed matter systems.
\par We propose here a novel mechanism involving Voids,
which can, in principle give rise to a difference between
the gravitational and the inertial mass, and
thereby, create effects similar to the dark matter.

\end{abstract}
\newpage

Basic  forces  of  nature  are  characterized by the existence of
fundamental constants.These constants not only  characterize  the
distance  scale  of motion of bodies in force field, it also sets
the  size  of  the  structure  that  can  arise. Eletromagnetism
probably  gives rise to the largest variety of structures. This is
because the scale in this case is set not just by the fine  structure
constant,  but  by  the  combination  of  fine  structure constant
$\alpha$ and various atomic numbers $Z$. The  strong  interaction
is  largely responsible for the existence of elementary particles
and nuclei. The weak interaction is responsible for the  decay  of
such    particle    and,    therefore,    sets    the   stability
standards. Gravitation, the the oldest known, interaction  is  not
as  well  understood as the other three fundamental interactions.
In the smallest scale we have the black hole singularity (distance
of  order  of  plank  length  $ ~\approx~  10^{-33} ~$  cm.),  in   the
intermediate  scale  the  problem  of  dark  matter and structure
formation (distance scale ranging
from a few kilopersec to a few megapersec) and in  the
largest  scale (distance scale of around  a few  hundred  megapersec),
where we have the expansion of universe, there is
the horizon problem and the problem of overall density, and also
the possibility that the universe may be accelarating in the present
epoch. In all these cases the constant that
enters in the description is the Newton's  constant  of
gravity,   $ ~G=6.67\times10^{-8}  ~cm^{3}/gm.sec^{2} ~$. The
ratio of the distance scales involved is approximately
$\approx ~10^{28}/10^{-33}=10^{61}~$,  where
as  in the case of eletromagnetism it is roughly the ratio of the
size   of   the   Jupiter   to   the   atomic   size   which   is
approximately $\approx
10^{10}/10^{-8}=10^{18}$,  in the case of strong interaction it
is $ ~\approx 10^{3} ~$.
Keeping in mind that there  is no  quantum  theory  of gravity as yet,
one wonders whether gravity is an effective theory with different
coupling constants governing motions at different  length  scales
(like the pion-nucleon coupling constant in strong interactions).
After  all,  the  direct verification of  gravity extends
in larger side of the scale
only up to distances $ ~\approx 10^{19} ~$ cms.  (a  few  parsecs  in
open clusters) and in smaller end of the scale not even submillimeter
level. This is the enigma of gravity. In this paper, we will be concerned
with only one of these problems.
We leave aside the problem of the smallest
and  largest  scale
and  consider only  the  intermediate
scale  (  from  a  few  kilopersec  to a few megapersec ). In this
distance scale, we have  the  problem  of the  dark  matter  and  the
structure  formation. {\bf They both probably are two sides of the same
coin.}

The  dark matter problem has been known for a long time. However,
in seventies and early eighties  when  old  data  was  analysed
properly  \cite{Einasto,Ostriker}  and newly obtained data was
analysed
\cite{Rubin} that the problem of dark matter was  well
established. We  expect
that the velocities of the satellite galaxies and large gaseous
clouds  orbiting around the galaxies should fall off with distance
as per the Coulomb law. However,  what  was  found  is  that  the
velocities  do  not decrease at all. They rather become constant,
and hence, the name flat rotation curves.
It is also found that in galactic clusters the velocities of
the galaxies are anomalously large and it is not possible to
explain it by the gravitational field of the luminous matter alone.
All  sorts  of  exotic
particles,  ditributed  around the galaxies with density $\rho(r)
\sim 1/r^{2}$ have been postulated to explain the flat  rotation
curves. In  these  models,  the  amount  of  dark matter needed to
explain the rotation curves is about eighty to ninety percent  of
total  mass of the galaxies. Therefore, for the formation of large
scale structure in the universe \cite{Padmanabhan,Peebles},  they
become  the  most  relevant  objects  and  dynamics  would almost
entirely be governed by them. All attempts to identify such  exotic
particles  have been elusive. However, the search continues. There
have also been search for the so called  MACHO  objects  ( baryonic
structures  of the size of Jupiter) which could also explain the
rotation curves. But  this  has  also  not  produced  satisfactory
results. If the dark matter is non-baryonic,
one would expect that the existence of vast amount of dark matter
would lead to the formation of some dark galaxies or clusters which
could be seen by gravitational lensing events. This could serve as a direct
evidence for dark matter. However, so far, there is no conclusive lensing
event pertaining to that.
On  the  other  hand,  the absence of dark matter would
imply that the theory of gravity be different at the intermediate
scale. At this stage there is  a  serious  psychological  barrier
based  on  very  genuine reasons. After all, Newtonian gravity is
the low velocity limit of the general theory of relativity.
There may exist data not explained by these  theories  but
there  is  no single  data  worth  mentioning which directly
contradicts them. Nevertheles, it does sound logical to  construct
models  and  look  for  their obvervational consequences. It is not
enough to construct models which only explains the flat  rotation
curves but, as mentioned before, it should also account for
the large scale  structure in the universe.

There  have  been  some  attempts to introduce such modifications.
Sanders  \cite{Sanders}  suggested  that  Newton's  constant   of
gravity  varies  with distance.The gravitational potential in his
model is

\begin{eqnarray}
U=-\frac{G(1-\beta e^{-r/r_0})}{r(1-\beta)}
\end{eqnarray}
The constants $\beta$ and $r_{0}$ can be
determined from the data.
However, there is no estimate for the size of
structures or its role in
large scale structure formation.
\par Milgrom and Bekenstein
\cite{Milgrom,Bekenstein}
assume that the ratio of inertial
mass $ ~m_{i} ~$ and the gravitational
mass $ ~m_{g} ~$ depends on accelaration
in the small accelaration limit. To be more precise
\begin{eqnarray}
m_{i}/m_{g} =\mu(a/a_{o})
\end{eqnarray}
where $ ~a ~$ is the accelaration and $ ~a_{o} ~$
is a universal constant,
$a_{o}=8h^{-2}\times10^{-8}cm.sec^{-2}$.
\begin{eqnarray}
\mu(a/a_{o})=1 \mbox{~for~} a\gg a_{o}
\nonumber \\
\mu(a/a_{o})=a \mbox{~for~} a\ll a_{o}
\end{eqnarray}
As  a  result  of this assumption the law of gravity becomes very
different when the  potential  gradient  is  small. Essentially, it
means, that  $ ~m_{i}a=m_{g}(GM/r^{2}) ~$ implies
$ ~m_{g}a^{2}=m_{g}(GM/r^{2}) ~$ in the small accelaration  limit. Now,
for a  rotating  body  $ ~a=v^{2}/r ~$  where  $v$  is  the  rotational
velocity.  The  above  relations  imply  that $ ~v^{4}=GM ~$,   and,
therefore,  the  rotational velocity does not depend on distance.
In this model, the kinetic energy is  too  soft  and  we  do  not
expect large structure to emerge. This mechanism  also fails to
accomodate an well established observation that
fainter (smaller) galaxies contain more dark matter.

\par Our  approach,  though  different,  accomodates some essential
ideas of Milgrom and Bekenstein. In their approach, they try to modify
the law of inertia for small accelarations. We, instead, look for
mechanism which can give rise to a  difference
between gravitational mass and the inertial
mass. Our approach is in the spirit of condensed matter physics. It is
in this sense that we view dark matter as an emergent phenomenon.
We shall argue in this paper that Voids \cite{zeldovich}
can, in principle,
create a difference between the gravitational and inertial mass. But
before we come to that, the following discussion is necessary. It is well
known that masses of particles can vary depending on the medium
it is moving in. In condensed matter physics this is very well known.
In fact, this is one of the basic ingredients in Landau Fermi
liquid theory \cite{abrikosov,pines1} .
One often considers variation of mass as an artefact
of quantum theory. But strictly speaking this is not correct
- though quantum theory is the essential tool for carrying out
analysis and quantitative calculations. The variation in
masses, essentially
reflects the nature of force field present in the medium. If the mass,
$m_{B}$, ($\mu= \frac{e\hbar}{2m_{B}c}$) ,
defining
the magnetic moment of an electron,
and the mass,
$m_{F}$, ($\epsilon_{F} =\frac{p^{2}}{2m_{F}}$),
defining the Fermi energy,
were always
equal, there would not be any diamagnetic substance in nature. The very fact
that there exist strong diamagnet such as bismuth, implies that
in these materials the "two masses" are different, $m_{F} < m_{B}$.
In such systems the inertial mass is lowered.
There are other materials known as heavy fermion systems ($CeAl_{3}$,
$CeCu_{6}$, $UPt_{3}$ ), in which the inertial mass increases - the
electrons acquire masses of several GeV.
The inertial masses of particles are modified by the force field in the
medium. Such a phenomenon is also known in
classical Hydrodynamics \cite{pines2}. The
increase or decrease of inertial mass is determined
by the nature of the drag the
medium exerts on moving bodies - if the drag is positive the mass increases
but if the drag is negative (as in the case of diamagnetic materials for
electrons), the inertial mass decreases. It is this idea, which we want
to carry over to the large structure. This requires review of some aspects
of standard cosmology and the theory of structure formation
at the large scale. This is briefly described below.
\par It is very well established through observation that
at distance scales of the order of 200-300 megapersec, the Universe is
isotropic and homogenous. This means that if we pick up a region of
the Universe of dimension 200-300 megapersec at any distance and in any
direction, it will contain the same amount of matter. Therefore, at this
scale the density of matter can be considered to be constant.
{\bf In Newtonian
gravity, such a distribution
of matter implies that at every point in space
the potential and force are unbounded \cite{landau} }.
This dilemma is resolved in the General
Theory of Relativity. For an isotropic and homogenous
distribution of matter one assumes the
Friedman-Robertson-Walker metric, given by the line element
\begin{eqnarray}
ds^{2}= dt^{2}-a^{2}(t)(\frac{dr^{2}}{1-kr^{2}}+r^{2}d\theta^{2}+r^{2}\sin^{2} \theta d\phi^{2})
\end{eqnarray}
in which the Einstein equation,
\begin{eqnarray}
R_{\mu \nu}-\frac{1}{2} g_{\mu \nu}R=\frac{8\pi G}{3} T_{\mu \nu}
\end{eqnarray}
takes the simple  form \cite{landau},
\begin{eqnarray}
\frac{\dot{a}^{2} +ka^{2}}{a^{4}} = \frac{8\pi G}{3} \rho_{0}
\end{eqnarray}
where $a(t)$ is the scale factor, $\rho_{0}$ is the averaged constant
density, and $k=1, -1$ or $0$, respectively for closed, open and  flat
universe.
This equation along with the equation of state describes
the isotropic and homogenous universe. The equations clearly show
that the isotropic and homogenous
distribution of matter can not be stable - the Universe
expands. The constant density serves as a source term in the
evolution of the scale factor. What is the source of gravity
in the large scale? When the mean free path of the particles
is small, matter can be treated as an ideal fluid and the Newton's
equations governing the motion of gravitating collisionless particles
in an expanding Universe can be written in terms of
$ ~{\bf x}= {\bf r}/a ~$ (the comoving space coordinate),
${\bf v} ={\bf \dot{r}}-H{\bf r}=a{\bf \dot{x}}$ (the peculiar velocity
field, H is the Hubble constant),
$\phi({\bf x},t)$ (the Newton gravitational potential) and
$\rho({\bf x},t)$ (the matter density). This give us the following
set of equations \cite{varun,strauss}.
Firstly, the Euler equation,
\begin{eqnarray}
\frac{\partial (a {\bf v})}{\partial t}+({\bf v.\nabla_{x}}){\bf v}=
-\frac{1}{\rho}{\bf \nabla_{x}} P-{\bf \nabla_{x}}\phi
\end{eqnarray}
Next, the continuity equation
\begin{eqnarray}
\frac{\partial \rho}{\partial t}+ 3H\rho +\frac{1}{a}{\bf \nabla_{x}}
(\rho {\bf v}) =0
\end{eqnarray}
And, finally the Poisson equation
\begin{eqnarray}
{\bf \nabla_{x}}^{2} \phi =4\pi Ga^{2}(\rho-\rho_{0})
=4\pi Ga^{2} \rho_{0} \delta
\end{eqnarray}
where $\rho_{0}$ is the mean background density and
$\delta=\rho/\rho_{0}-1$ is the density contrast.

Therefore, at large scale, the source of gravity is not
the average density $\rho_{0}$ but
the density fluctuations, $\delta \rho >0$.
It is a subject of study in theory of structure formation as to what
kind of density fluctuation would grow in time and lead to the formation of
galaxies, and clusters and superclusters of galaxies
\cite{Padmanabhan,Peebles,varun,strauss}.
It is important, here, to remember that at the scale of dimensions,
200 - 300 megapersec, the Universe is homogenous and isotropic
and acquires constant density and, therefore, if in some subregion
$\delta \rho >0$, there must be some subregion where
$\delta \rho <0$, so as to
reproduce the constant density profile. These domains
with $\delta \rho <0$ are known as {\bf Voids} \cite{zeldovich,varun}.
Note that for Voids $ ~\delta\rho/\rho_{0} ~$ is always bounded below
by $-1$ . Such regions of Voids
dominate the volume in the universe giving rise to cellular structures
with the clusters and superclusters of galaxies forming string like walls
around them. Existence of Voids are supported by direct observation as well
as numerical simulation of hydrodynamic equations
\cite{varun,ryden,hoyle,antonu}.
The observed Voids seem to have dimension of several (tens of)
megapersecs. However, Voids of our interest would be those whose dimensions
are a few (tens of) kilopersec distributed within the intergalactic medium
and within clusters and supercluetrs of galaxies. There is, so far,
no observational evidence for Voids of such dimensions.
However, issue of how the Voids are distributed is not yet settled. There
are some numerical simmulations which suggest that Voids may form
a connected network (\cite{varun} and references there in). This connectivity
of Void network is very important for cosmology. If the universe starts with
a {\bf matter coonnected network} and latter
makes a transition to {\bf Void connected
network}, the rate of expansion will undergo abrupt change.
\par Let us suppose that Voids of such small dimensions exist and are
distributed in the intergalactic medium and the clusters and superclusters
of galaxies. The question that immediately arises is related to the observation
of such objects. What are the observational signature of such Voids? Large
Voids are easily identified by the absence of galaxies and clusters in those
regions.For smaller size Voids, possibly the ultra-low density limit of
Saha ionization formula could be of some help because the Voids
are ultra-low density regions. The formula implies that the atoms trapped
in the Voids would prefer to remain ionized, and therefore, the random
motion of ions and electrons would give rise to a faint radiation glow (see
appendix for details).
\par Let us proceed with the assumption that Voids as discussed above exists.
Since the source term responsible for the formation of Voids is
the negative density contrast, it will give rise to force field that is
opposite to that of gravity. Therefore, the intergalactic
medium will be more like a {\bf polarised region} - where
Galaxies and Voids
give rise to two opposite type of force fields.What effect
would the Void
have on a massive body? It would clearly exert a repulsive force
at least locally.
Therefore, a massive body moving around a galaxy or in the
clusters would feel
an extra push from the Voids. This effect can be taken into account by
assuming that the inertial mass is effectively decreased.
Thus the effect of Voids
would brings about a difference in the "effective" inertial mass and the
gravitational mass. This difference would apper as if there is
some dark matter.
\par I would like to thank Prof. Ashok Sen, Harischandra Research Institute,
Allahabad, U.P., India and Prof. D.Narashimha, Tata Institute of
Fundamental Reaserch, Mumbai, India for discussion and suggestion in
the early phase of the work.
\newpage
\centerline{ \bf{APPENDIX} }
\vskip .1 in
\centerline{Ultra-low desity limit of the Saha ionization formula}
 The Saha ionization formula \cite{saha} has played a very impotant
 role in the development of astrophysics. The ultra-low density limit
 of this formula has been known for a long time \cite{feynman}. In
 this limit, the formula suggests that the atoms in equillibrium
 prefer to remain  in ionized state.This ionization, just from
 "expansion" as the density goes down, has been listed as one
 of the surprises in theoretical physics by Peierls \cite{peierls}.
 In this appendix, we point out that this
 ultra-low density limit of Saha ionization formula is very relevant for
 Voids.

 The ionization formula is given by,
 \begin{eqnarray}
 \frac{n_{e}n_{i}}{n_a}=\frac{1}{v_a}e^{-W/kT}
 \end{eqnarray}
 In the equation above, $n_e$ , $n_i$, $n_a$ are the densities of
 electrons, ions and atoms(not ionized) respectively. $W$ is the
 ionization potential, $T$ is the temperature and $k$ is the Boltzman
 constant.The volume occupied by a bound electron at
 temperature $T$ is represented by $v_a$.
 It is, essentially, the
 volume contained within a thermal de Brogle wave length.
 \begin{eqnarray}
 v_a=\lambda_{th}~^{3}=\Big(~\frac{2\pi \hbar^2}{m_{e}kT}~\Big)^{3/2}
 \end{eqnarray}
 Let us consider a box of volume $V$ which, to start with,
 contains N number of hydrogen atoms. Let a fraction $X$ of them
 be ionized. In this case , $~n_e~=~\frac{N}{V}~X~=~n_i~$ and
 $~n_a~=~(1-X)~\frac{N}{V}~$. Substituting these values
 in the ionization formula we obtain,
 \begin{eqnarray}
 \frac{X^2}{1-X}~\frac{N}{V}~=~\frac{1}{v_a}e^{-W/kT}
\end{eqnarray}
 From the equation above, we see that the fraction of
 charged particles in equilibrium increases when we increase
 the volume(i.e., decrease the density). In the ultra-low density
 ($~\frac{N}{V}~ \rightarrow~~ 0$ or
 $~\frac{V}{N}~ \rightarrow~~\infty$ ), atoms would prefer to
 remain ionized .

\par The Voids are the ultra-low density regions in the Universe,
 and these are the regions where one would expect to observe
 the consequences of ultra-low density limit of the Saha ionization
 formula. As discussed before, in this limit, atoms would prefer to
 remain ionized.
Saha ionization  formula implies that, at ultra-low density,
once the ionization takes place there is hardly any chance for
recombination \cite{feynman,peierls,ghosh}. Therefore, the random motion
of the charged particles in the Voids should create a faint radiation glow.
\par At this stage, the important question is: what is the
 source of the ionization energy? One common source of ionization
 are the starlights. However, at high
 red shift, their intensity is very low.
 The most common source of ionization
 energy at high red shift are the lights from Quasars.


\begin{thebibliography}{99}
\bibitem{Einasto}J.Einasto, A.Kaasik and E.Saar-Nature,{\bf 250} 309(1974)
\bibitem{Ostriker}J.P. Ostriker, P.J.E.Peebles and A.Yahil- Astrophys.J
           Lett.,{\bf 193} L1(1974)
\bibitem{Rubin}V.C.Rubin- Science,{\bf 220} 1339 (1983)
\bibitem{Padmanabhan} T.Padmanabhan- Structure Formation in the Universe,
           Cambridge, Cambridge University Press, 1993
\bibitem{Peebles} P.J.E.Peebles-The Large Scale Structure in the Universe,
           Princeton University Press, Princeton, 1980
\bibitem{Sanders} R.H.Sanders-Astronom.Astrophys.,{\bf 136} L21(1984)
\bibitem{Milgrom} M.Milgrom-Astrophys.J.,{\bf 270} 365(1983)
\bibitem{Bekenstein} J.Bekenstein and M.Milgrom-Astrophys.J.,
{\bf 286} 7(1984)
\bibitem{zeldovich} Ya.B.Zeldovich,J.Einasto and S.F.Sandarin-Nature
{\bf 300} 407(1982)
\bibitem{abrikosov} A.A.Abrikosov,L.P.Gorkov and I.E.Dzyaloshinski-
 Methods of of Quantum Field Theory in Statistical Physics, Prentice-Hall,
 INC,Englewood Cliffs, New Jersey, 1963
\bibitem{pines1} P.Nozieres and D.Pines-The Quantum Theory of Fluids-
 {\bf Vol-1}, Addison-Wisely Publishing Company, INC.,1990
\bibitem{pines2} P.Nozieres and D.Pines-The Quantum Theory of Fluids-
 {\bf Vol-2,Section-3.5,Page-43,Eq.(3.21)}, Addison-Wisely Publishing
 Company, INC.,1990
\bibitem{landau} L.D.Landau and E.M.Lifshitz-The Classical Theory of
 Fields, ({\bf 4th.Ed.}, 1975, Pergamon Press Ltd., Oxford ,England),
           Chapter{\bf 12}
\bibitem{varun} V.Sahni and P.Coles -Physics Reports {\bf 262} 1(1995)
                (Section 5)
\bibitem{strauss} M.A.Strauss and J.A.Willick-Physics Reports {\bf 261}
271(1995)
\bibitem{ryden} B.S.Ryden and A.L.Mellot- APJ {\bf 470} 160(1996)
\bibitem{hoyle} F.Hoyle and M.S.Vogeley- arXiv:astro-ph/0109357
\bibitem{antonu} V.Antonuccio-Delogu et.al.,-Mon.Not.R.Soc.{\bf 000}
1(2000)
\bibitem{saha} Saha,M.N. 1920,Phil.Mag.{\bf 40} 472
\bibitem{feynman} Feynman,R.P., Leighton,R.
and Sands,M. 1963 The Feynman Lectures on Physics, Chapter {\bf 42},
Section {\bf 42.3},The Addison-Wisely Publishing Inc
\bibitem{peierls} Peierls,R. 1979,Surprises in Theoretical
Physics,Princeton University Press, Princeton, NJ,  {\bf pp 52-5},
\bibitem{ghosh} Ghosh,K. and Ghosh,G. 1998,
Eur.J.Phys. {\bf 19} 7


\end{thebibliography}
\end{document}